\documentclass[aps,showpacs,superscriptaddress,nofootinbib,preprint]{revtex4}
\usepackage{graphicx}
\def\slashchar#1{\setbox0=\hbox{$#1$}
   \dimen0=\wd0 \setbox1=\hbox{/} \dimen1=\wd1
   \ifdim\dimen0>\dimen1 \rlap{\hbox to \dimen0{\hfil/\hfil}} #1
   \else  \rlap{\hbox to \dimen1{\hfil$#1$\hfil}} / \fi}
\def\p{\slashchar{p}}
\def\q{\slashchar{q}}
\def\D{\slashchar{D}}

\begin{document}
\title{Charged lepton induced one kaon production off the nucleon}

\author{M. Rafi \surname{Alam}}
\affiliation{Department of Physics, Aligarh Muslim University, Aligarh-202 002, India}
\author{I. \surname{Ruiz Simo}}
\affiliation{Dipartimento di Fisica, Universit\`a degli studi di Trento,
I-38123 Trento, Italy}
\affiliation{Departamento de F\'\i sica At\'omica, Molecular y Nuclear,
Universidad de Granada, E-18071 Granada, Spain}
\author{M. Sajjad \surname{Athar}}
\affiliation{Department of Physics, Aligarh Muslim University, Aligarh-202 002, India}
\author{M. J.  \surname{Vicente Vacas}}
\affiliation{Departamento de F\'\i sica Te\'orica and IFIC, Centro Mixto
Universidad de Valencia-CSIC, Institutos de Investigaci\'on de
Paterna, E-46071 Valencia, Spain}

\begin{abstract}
We study single kaon production off the nucleon induced by electrons (positrons) 
 i.e. $e^-(e^+) + N \rightarrow \nu_e (\bar \nu_e) + \bar K (K) + N^\prime$ at low energies.
 The possibility of observing these processes with the high luminosity 
beams available at TJNAF and Mainz  is discussed, taking into account that the strangeness conserving electromagnetic reactions have 
a higher energy threshold for $\bar K (K)$ production.
 The calculations are done using a microscopic model that 
starts from the SU(3) chiral Lagrangians, includes background terms 
and the resonant mechanisms associated to the lowest lying resonance $\Sigma^*(1385)$.
\end{abstract}
\pacs{12.15.-y,13.60.Le,25.30.Hm,25.30.Rw}

\maketitle
\section{Introduction}
Recently the importance of the study of kaon production induced by real 
and virtual photons on nucleons and nuclei has been emphasized 
due to the development of accelerators like MAMI, JLAB, LNS, ELSA, SPring-8, GRAAL, 
etc.~\cite{Achenbach:2011eb,Achenbach:2011zza,Achenbach:2012zz,Beckford:2012rb,Zegers:2003ux,Sumihama:2005mt,Lleres:2007tx,Lawall:2005np,Glander:2003jw,Carman:2002se,Ambrozewicz:2006zj}.  
In particular, the availability of very high luminosity beams has provided the opportunity to 
study the electromagnetic associated strangeness 
production~\cite{Mart:1995wu,Mart:2012zz,Li:1995si,Steininger:1996xw,Borasoy:2007ku,Kaiser:1996js,
Feuster:1998cj,Corthals:2005ce,Anisovich:2005tf,JuliaDiaz:2006is,Guidal:1997hy} 
of a strange and an anti-strange particle. 

 The cross section for weak associated strangeness production is obviously much smaller 
than that of the electromagnetic one. However, weak interaction allows for processes where only 
one strange/anti-strange particle is produced, ($\Delta S=\pm 1$) and these reactions could have 
a substantially lower threshold.
 For instance, the threshold for electron induced weak $K^-$ production on a proton is around 
600 MeV whereas it is 1.5 GeV for electromagnetic production, as an additional kaon is required.

The study of these reactions could provide valuable information on the coupling constants $D$ and $F$ 
that govern the interaction of the SU(3) lightest baryon octet with the pseudoscalar mesons and 
also their $\beta$ decays. 
More specifically, the  $g_A(=D+F)$  combination, related to the neutron $\beta$ decay, is very well known,
 but the knowledge of the $D$ and $F$ values is less precise \cite{Cabibbo:2003cu}.
 Also, one may investigate the $Q^2$ dependence of the weak axial form factors of nucleons and hyperons. 
In this paper, we explore the possibility of doing such experiments and present a quantitative 
analysis of the charged current reaction in which a kaon/antikaon is produced without conserving $\Delta$S.
This study is based on our earlier works on neutrino/antineutrino induced single kaon production~\cite{Alam:2012zz,RafiAlam:2010kf,AlvarezRuso:2012fc}
and the same formalism is applied here to study one kaon production off the nucleon obtained from electron as well as positron beams.

We proceed by introducing the formalism in brief in Sec.~\ref{sec:form}. Results and discussions are presented in
Sec.~\ref{sec:res}.

\section{Formalism}
\label{sec:form}
The processes considered here are the charged lepton induced weak $| \Delta S | = 1$ $K(\bar{K})$ production. 
The  single antikaon production channels induced by electrons are
\begin{eqnarray}\label{eq:elec}
 e^- + n \rightarrow \nu_e + K^- + n, \;\; e^- + p \rightarrow \nu_e + \bar{K^0} + n, \;\; e^- + p \rightarrow \nu_e + K^- + p,  
 \end{eqnarray}
and the corresponding positron induced channels are
\begin{eqnarray}\label{eq:posi}
e^+ + n \rightarrow \bar \nu_e + K^+ + n, \;\; e^+ + n \rightarrow \bar \nu_e + K^0 + p, \;\; e^+ + p \rightarrow \bar \nu_e + K^+ + p 
\end{eqnarray}

The expression for the differential cross section in the laboratory frame for the above processes is given by
\begin{eqnarray}\label{sigma_inelas}
d^{9}\sigma &=& \frac{1}{4 M E_e(2\pi)^{5}} \frac{d{\vec k}^{\prime}}{ (2 E_{\nu})} 
\frac{d{\vec p\,}^{\prime}}{ (2 E^{\prime}_{p})} \frac{d{\vec p}_{k}}{ (2 E_{k})}
 \delta^{4}(k+p-k^{\prime}-p^{\prime}-p_{k})\bar\Sigma\Sigma | \mathcal M |^2,
\end{eqnarray}
where  $ k( k^\prime) $ is the momentum of the incoming(outgoing) lepton with energy $E_e( E_\nu)$, 
$p( p^\prime)$ is the momentum of the incoming(outgoing)
nucleon with mass $M$. The kaon 3-momentum is $\vec{p}_k $ having energy $ E_k  $,
$ \bar\Sigma\Sigma | \mathcal M |^2  $ is the square of the transition amplitude
 averaged(summed) over the spins of the initial(final) state. The transition amplitude may be written as 
\begin{equation}
\label{eq:Gg}
 \mathcal M = \frac{G_F}{\sqrt{2}}\, j_\mu J^{\mu}=\frac{g}{2\sqrt{2}}j_\mu \frac{1}{M_W^2}
\frac{g}{2\sqrt{2}}J^{\mu},
\end{equation}
 where $j_\mu$ and $  J^{\mu}$ are the leptonic and hadronic currents respectively, 
$G_F=\sqrt{2} \frac{g^2}{8 M^2_W}$ is the Fermi coupling constant, 
$g$ is the gauge coupling and $M_W$ is the mass of the $W$-boson.

First, we shall discuss the leptonic current, the hadronic current and the transition amplitude
corresponding to the reactions shown in 
Eq.~\ref{eq:elec}. The leptonic current is obtained 
from the Standard Model Lagrangian coupling of the $W$-boson to the leptons 
\begin{equation}\label{eq:Lag}
{\cal L}=-\frac{g}{2\sqrt{2}}\left[{ W}^-_\mu\bar{l}\gamma^\mu (1-\gamma_5)\nu_l  + 
{ W}^+_\mu\bar{\nu}_l \gamma^\mu(1-\gamma_5)l \right] 
=-\frac{g}{2\sqrt{2}}\left[j^\mu_{(L)}{ W}^-_\mu+h.c.\right].
\end{equation}

\begin{figure}
\begin{center}
\includegraphics[width=0.8\textwidth,height=.4\textwidth]{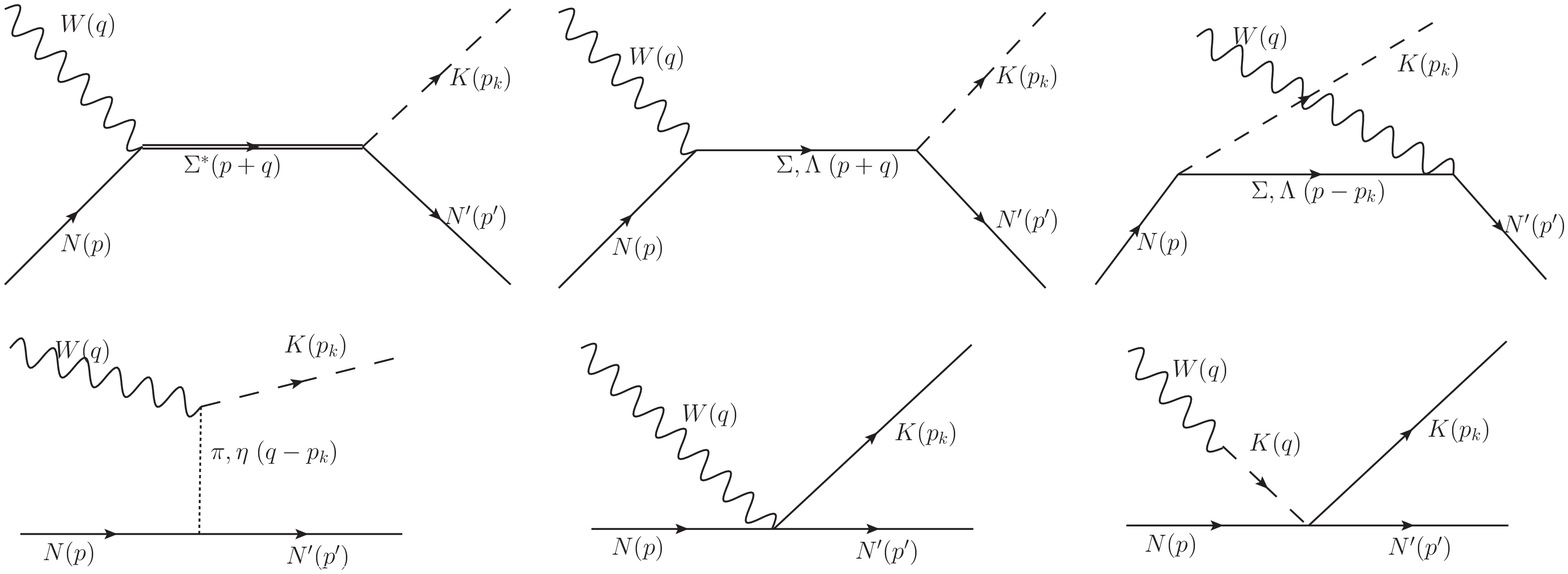}
\caption{Feynman diagrams for the processes $e^-N\rightarrow \nu_e N^\prime \bar K$ and $e^+N\rightarrow \bar \nu_e N^\prime K $. 
Here $\bar K$ stands for a $K^-$ or $\bar K^0$ obtained in an electron induced process and K stands for $K^+$ or $K^0$ obtained
in a positron induced process.
First row 
from left to right: s-channel $\Sigma^*$ resonance term (labeled SCR in the text) ,  s-channel (SC) and u-channel (UC)
 $\Sigma,\Lambda $ propagator; second row:  Pion/Eta meson ($ \pi P/ \eta P $)  exchange terms,
Contact term (CT) and finally kaon pole term (KP).}
\label{fg:terms}
\end{center}
\end{figure}

The Feynman diagrams that  contribute to the hadronic current are depicted in Fig.~\ref{fg:terms}.
There is a meson ($\pi$P,$\eta$P) exchange term, a contact term (CT) and a kaon pole (KP) term.
 For the electron induced reactions we also have the s-channel diagrams with 
$\Sigma,\Lambda$(SC) and $\Sigma^*$(SCR) as intermediate states.  
In the case of positron induced reactions the s-channel diagrams 
(SC and SCR) do not contribute, but we must include the u-channel (UC) one.
The contributions to the hadronic current coming from different terms are 
written using the Lagrangian obtained from chiral perturbation theory($\chi$PT).
The lowest-order SU(3) chiral Lagrangian describing the interaction of pseudoscalar mesons in the 
presence of an external weak current is written as~\cite{Scherer:2002tk}:
\begin{equation}
\label{eq:lagM}
{\cal L}_M^{(2)}=\frac{f_\pi^2}{4}\mbox{Tr}[D_\mu U (D^\mu U)^\dagger]
+\frac{f_\pi^2}{4}\mbox{Tr}(\chi U^\dagger + U\chi^\dagger),
\end{equation}
where $f_\pi$(=92.4MeV) is the pion  decay constant, $U(x)=\exp\left(i\frac{\phi(x)}{f_\pi}\right)$ 
is the SU(3) representation of the pseudoscalar  meson octet fields $\phi(x)$
and $D_\mu U$ is its covariant derivative given by:
\begin{eqnarray}
D_\mu U&\equiv&\partial_\mu U -i r_\mu U+iU l_\mu\,.
\end{eqnarray}
Here, $l_\mu$ and $r_\mu$ correspond to left and right handed currents, that for the charged current(CC) case are given by
\begin{equation}
r_\mu=0,\quad l_\mu=-\frac{g}{\sqrt{2}}
({W}^+_\mu T_+ + {W}^-_\mu T_-),
\end{equation}
with $W^\pm$, the $W$ boson fields. The  elements of the 
Cabibbo-Kobayashi-Maskawa matrix $V_{ij}$ (for light quarks) can be written as 
$$
T_+=\left(\begin{array}{rrr}0&V_{ud}&V_{us}\\0&0&0\\0&0&0\end{array}\right);\quad
T_-=\left(\begin{array}{rrr}0&0&0\\V_{ud}&0&0\\V_{us}&0&0\end{array}\right),
$$

The lowest order baryon-meson interaction Lagrangian coupling to the W-boson field is given by:
\begin{equation}\label{eq:lagB}
{\cal L}^{(1)}_{MB}=\mbox{Tr}\left[\bar{B}\left(i\D
-M\right)B\right]
-\frac{D}{2}\mbox{Tr}\left(\bar{B}\gamma^\mu\gamma_5\{u_\mu,B\}\right)
-\frac{F}{2}\mbox{Tr}\left(\bar{B}\gamma^\mu\gamma_5[u_\mu,B]\right),
\end{equation}
where $M$ denotes the mass of the baryon octet $B$. For the coupling constants we take the values  
$D=0.80$ and $F=0.46$ which have been determined from the baryon semileptonic decays~\cite{Cabibbo:2003cu}. 

The covariant derivative of $B$ is given by
\begin{equation}
D_\mu B=\partial_\mu B +[\Gamma_\mu,B],
\end{equation}
with
\begin{equation}
\Gamma_\mu=\frac{1}{2}\left[u^\dagger(\partial_\mu-ir_\mu)u
+u(\partial_\mu-il_\mu)u^\dagger\right],
\end{equation}
where  we have introduced a new variable $u$, $u^2=U$. Finally, 
\begin{equation}
u_\mu= i\left[u^\dagger(\partial_\mu-i r_\mu)u-u(\partial_\mu-i
l_\mu)u^\dagger\right].
\end{equation}

We have also included  the contribution of terms with the $\Sigma^*(1385)$ resonance belonging to the SU(3) baryon decuplet,
 which is near the threshold of the $NK$ system. 
 This is  suggested by the dominant role played by the $\Delta(1232)$ in pion production reactions.
For the weak excitation of the $\Sigma^*(1385)$ resonance and its subsequent decay in $NK$, the lowest order 
SU(3) Lagrangian coupling the pseudoscalar mesons
with decuplet-octet baryons in presence of external weak current is given by:
\begin{equation}
{\cal L}_{dec} = {\cal C} \left( \epsilon^{abc} 
\bar T^\mu_{ade} u_{\mu,b}^d B_c^e + 
\, h.c. \right),
 \label{eq:dec_lag}
\end{equation}
where $T^\mu$ is the SU(3) representation of the decuplet fields, $a-e$ are flavor indices
\footnote{The physical states of the decuplet are: $T_{111} = \Delta^{++},T_{112} = \frac{\Delta^{+}}{\sqrt3},
T_{122} = \frac{\Delta^{0}}{\sqrt3}, T_{222} = \Delta^{-},T_{113} = \frac{\Sigma^{*+}}{\sqrt3},
T_{123} = \frac{\Sigma^{*0}}{\sqrt6},T_{223} = \frac{\Sigma^{*-}}{\sqrt3},T_{113} =  \frac{\Xi^{+}}{\sqrt3},
T_{133} =  \frac{\Xi^{0}}{\sqrt3},T_{333} = \Omega^{-} $.}.

The parameter
${\cal C}\simeq 1$  has been fitted to the $\Delta(1232)$  decay-width.  
The spin 3/2  propagator for
$\Sigma^*$  is given by
\begin{equation}
G^{\mu\nu}(P)= \frac{P^{\mu\nu}_{RS}(P)}{P^2-M_{\Sigma^*}^2+ i M_{\Sigma^*} \Gamma_{\Sigma^*}},
\qquad 
\end{equation}
where $P=p+q$ is the momentum carried by the resonance, $q=k-k^\prime$ and 
$P^{\mu \nu}_{RS}$ is the projection operator
\begin{equation}
P^{\mu\nu}_{RS}(P)= \sum_{spins} \psi^{\mu} \bar \psi^{\nu} =- (\slashchar{P} + M_{\Sigma^*}) \left [ g^{\mu\nu}-
  \frac13 \gamma^\mu\gamma^\nu-\frac23\frac{P^\mu
  P^\nu}{M_{\Sigma^*}^2}+ \frac13\frac{P^\mu
  \gamma^\nu-P^\nu \gamma^\mu }{M_{\Sigma^*}}\right],
\label{eq:rarita_prop}
\end{equation}
with $M_{\Sigma^*}$ the resonance mass  and $\psi^{\mu}$ 
 the Rarita-Schwinger spinor. The  $\Sigma^*$ width obtained using the
Lagrangian  of Eq.~\ref{eq:dec_lag}  can be written as
\begin{eqnarray}
 \Gamma_{\Sigma^*}&=&\Gamma_{\Sigma^*\rightarrow \Lambda \pi} 
+ \Gamma_{\Sigma^*\rightarrow \Sigma \pi}+ \Gamma_{\Sigma^*\rightarrow N \bar{K}}\; ,
\label{eq:width}
\end{eqnarray}
where
\begin{eqnarray}
 \Gamma_{\Sigma^* \rightarrow Y,\, meson}&=&\frac{C_Y}{192\pi}\left(\frac{\cal C}{f_\pi}\right)^2
\frac{(W+M_Y)^2-m^2}{W^5}\lambda^{3/2}(W^2,M_Y^2,m^2) \,
\Theta(W-M_Y-m).
\end{eqnarray}
Here, $m,\, M_Y$ are the masses of the emitted meson and baryon.
 $\lambda(x,y,z)=(x-y-z)^2-4yz$ and $\Theta$ is the unit step function. 
 The factor $C_Y$   is 1 for $\Lambda$ and $\frac23$ for $N $ and $\Sigma$. 

Using symmetry arguments, the most general $W^- N \rightarrow \Sigma^*$ vertex can be written in terms of 
a vector and an axial-vector part as,
\begin{eqnarray} \label{eq:delta_amp}
\langle \Sigma^{*}; P= p+q\, | V^\mu | N;
p \rangle &=& V_{us} \bar\psi_\alpha(\vec{P} ) \Gamma^{\alpha\mu}_V \left(p,q \right)
u(\vec{p}\,), \nonumber \\
\langle \Sigma^{*}; P= p+q\, | A^\mu | N;
p \rangle &=& V_{us} \bar \psi_\alpha(\vec{P} ) \Gamma^{\alpha\mu}_A \left(p,q \right)
u(\vec{p}\,),
\end{eqnarray}
where
\begin{eqnarray}
\Gamma^{\alpha\mu}_V (p,q) &=&
\left[ \frac{C_3^V}{M}\left(g^{\alpha\mu} \slashchar{q}-
q^\alpha\gamma^\mu\right) + \frac{C_4^V}{M^2} \left(g^{\alpha\mu}
q\cdot P- q^\alpha P^\mu\right)
+ \frac{C_5^V}{M^2} \left(g^{\alpha\mu}
q\cdot p- q^\alpha p^\mu\right) + C_6^V g^{\mu\alpha}
\right]\gamma_5 \nonumber\\
\Gamma^{\alpha\mu}_A (p,q) &=& \left [ \frac{C_3^A}{M}\left(g^{\alpha\mu} \slashchar{q}-
q^\alpha\gamma^\mu\right) + \frac{C^A_4}{M^2} \left(g^{\alpha\mu}
q\cdot P- q^\alpha P^\mu\right)
+ C_5^A g^{\alpha\mu} + \frac{C_6^A}{M^2} q^\mu q^\alpha
\right ]. \label{eq:del_ffs}
 \end{eqnarray}
The details of the $C_i$ N-$\Sigma^*$ transition form factors are given in Ref.~\cite{Alam:2012zz,Hernandez:2010bx}.
For all background terms,  we adopt a global dipole form factor
$
F(q^2)=1/(1-q^2/M_F^2)^2,
$ 
with a mass $M_F\simeq 1.05$ GeV that multiplies  the hadronic currents. 
Its effect, for energies of electron presently available at the accelerators will be discussed.
 
The final expressions of the hadronic currents $j^\mu$ for the electron as 
well as positron induced processes are given in the Appendix-~\ref{app:amplitude} and 
the various parameters of the currents are shown in Table~\ref{tb:currents}.

\section{Results and Discussion}\label{sec:res}
\begin{figure}
\begin{center}
 \includegraphics[height=11cm]{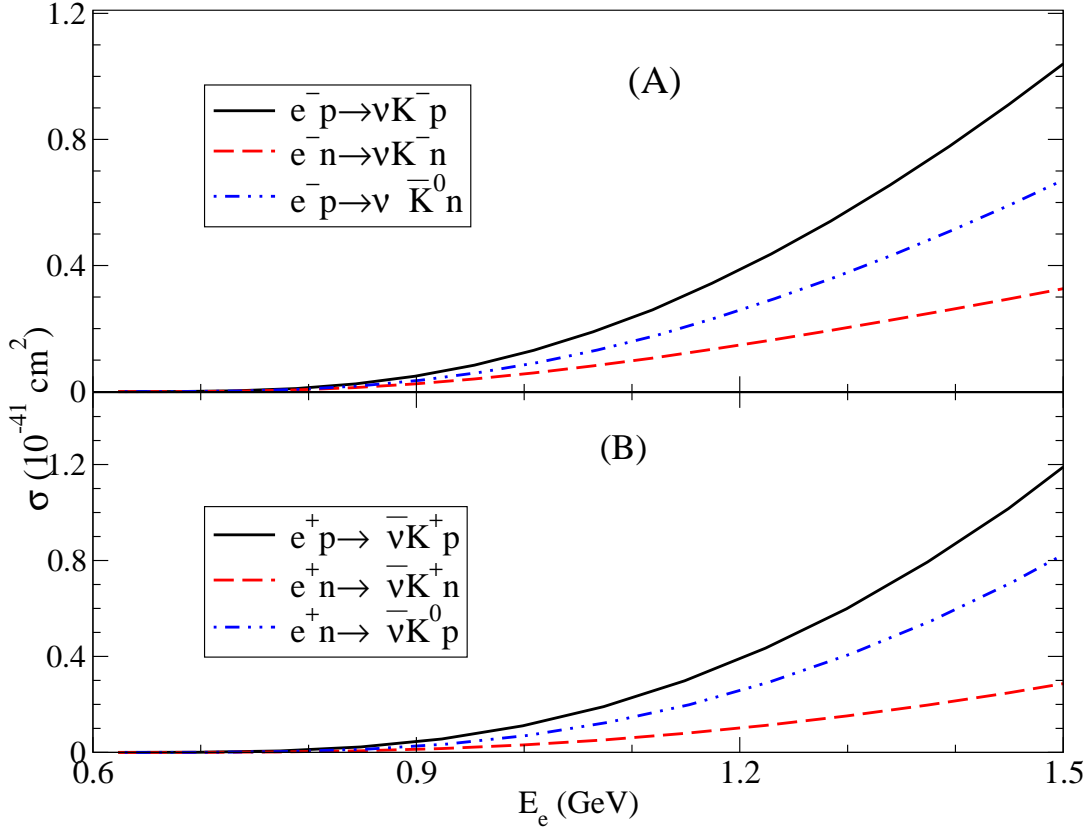}
\caption{Cross section $\sigma$ vs electron(positron) energy $E_e$ for the $\bar K (K) $ production}
\label{fg:xsec}
\end{center}
\end{figure}
Firstly, we present in Fig.~\ref{fg:xsec} the results for the total scattering cross section $\sigma$ for 
the reactions given in Eqs.~\ref{eq:elec} and \ref{eq:posi}. 

\begin{figure}
\begin{center}
\includegraphics[height=11cm]{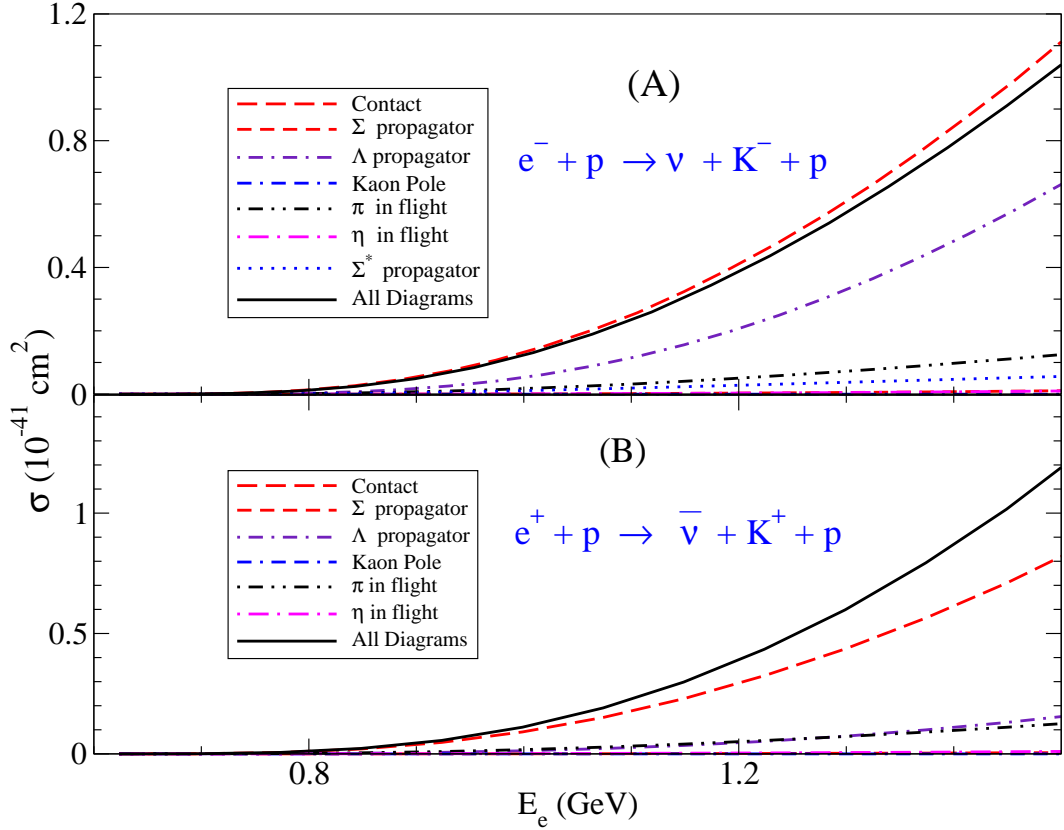}
\caption{Contribution of the different terms to the total cross section $\sigma$ vs electron(positron) 
energy $E_e$ for the $\bar K (K)$ production}
\label{fg:xsecpp}
\end{center} 
\end{figure}

We find that $ e^-(e^+) + p \rightarrow \nu_e(\bar\nu_e) + K^-(K^+) + p $
has the largest cross section followed by $ e^-(e^+) + p(n) \rightarrow \nu_e(\bar\nu_e) + \bar{K^0}(K^0) + n(p)$ and
 $ e^-(e^+) + n \rightarrow \nu_e(\bar\nu_e) + K^-(K^+) + n$. 
Furthermore, we find that the cross sections  for the positron induced processes are 
larger than for the corresponding electron induced process.
 This is basically due to the different interference between the s-channel  and  contact terms,
 as can be seen in Fig.~\ref{fg:xsecpp}, where we explicitly show the contribution of the individual terms of the hadronic current for two 
channels\footnote{Certainly, these individual contributions are not observable and they are shown here to help 
explaining the sensitivity of the physical processes to the various parameters.}.

We find that the contribution of the contact term is the largest followed by the mechanism with a
 $\Lambda$ in the  intermediate state, the $\pi$ pole term, etc. 
The  contribution due to the $\Sigma^\ast$ in cross section for the discussed energies is quite small, for example it is around 10\% at 
$E_e=$1 GeV and 5\% at $E_e=$1.5 GeV of the total cross section. 
Therefore, these reactions are not suitable to learn about the $\Sigma^\ast(1385)$ resonance properties.
The contributions of $\Lambda$ intermediate states both in UC and SC are larger than those corresponding to the $\Sigma$ hyperon,
 which can be easily understood by the corresponding  Clebsch-Gordan coefficients.
 Similar is the trend for the other channels not shown in the figure.

Given the smallness of the resulting cross sections, it is important to consider the feasibility of their experimental measurement. 
Here, we will only discuss electron processes. Let us remark that at energies below 1.5 GeV and in electron induced reactions, the 
presence of an antikaon in the final state fully defines the process, in the sense that it is necessarily a charge exchange weak 
production process and there is no other additional strange particle in the final 
state\footnote{Processes with additional particles, such as a pion, are also expected to be much smaller because of their phase space.}.
This is so because of the higher energy threshold of any other mechanism that could produce antikaons. Therefore, there is no need to 
measure other particles in coincidence.

To estimate the number of events for single kaon production we have considered a luminosity of 
 $5\times 10^{37}\, s^{-1}$ cm$^{-2}$ for MAMI, that corresponds to a 10 cm liquid hydrogen target at an electron beam 
 current of 20 $\mu$A as described in~\cite{Altarev:2005gp}. For TJNAF, we take a luminosity of
 $5\times 10^{38}\, s^{-1}$ cm$^{-2}$ that corresponds to a current of 100 $\mu$A  and a larger  
 liquid hydrogen target~\cite{Aniol:2004hp} that has been used on the measurement of parity 
 violating electron proton scattering. Under these conditions and for  1.5 GeV electrons, 
 we would have some 480 events per day for the reaction  $ e^- + p \rightarrow \nu_e + K^- + p $ at 
 TJNAF (48 at MAMI). For the $e^- + p \rightarrow  \nu_e + \bar{K^0} + n $ reaction, we would get 320 
 events per day at TJNAF (32 at MAMI).  Certainly, the numbers could be changed by using different 
 targets and/or current but equally important is the efficiency in the kaon detection, that  
 depends on the kaon kinematics and the detector. 

\begin{figure}
\begin{center}
\includegraphics[width=10cm]{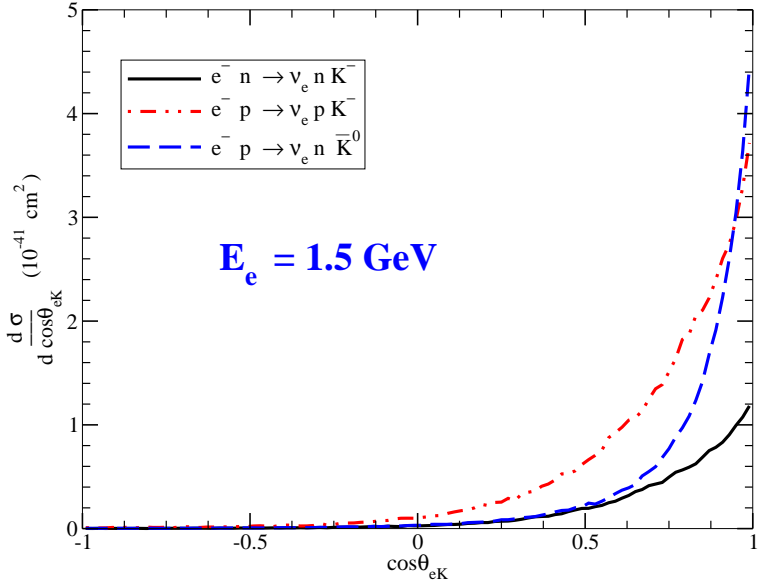}
\caption{ Kaon angle distributions at electron energy $E_e = 1.5$ GeV}
\label{fg:ang}
\end{center}
\end{figure}

\begin{figure}
\begin{center}
\includegraphics[width=10cm]{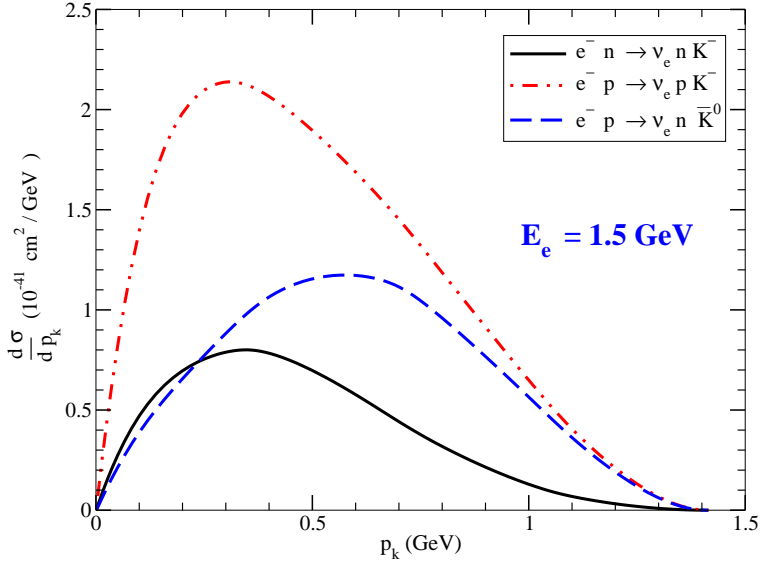}
\caption{Kaon momentum distributions at electron energy $E_e = 1.5$ GeV}
\label{fg:mom}
\end{center}
\end{figure}

The kaon angle and momentum distributions for the electron induced processes  are shown in Figs.~\ref{fg:ang}  
and~\ref{fg:mom} at an electron energy of E$_e= 1.5$ GeV,  
that could be appropriate for both the TJNAF and  MAMI facilities. 
As shown in Fig.~\ref{fg:ang}, the three channels under study are forward peaked, 
specially for the $\bar{K}^0$ production.
The momentum distributions peak around 0.3 GeV for $e^- n \rightarrow \nu_e  K^-  n, \; \text{and} \;   e^-  p \rightarrow \nu_e  K^- p $
while for $ e^-  p \rightarrow \nu_e  \bar{K^0}  n$ it peaks around 0.6 GeV.
This is because the contact term, 
which ̄has dominant contribution for $ K^- $ production channels, peaks at low $p_k$. 
For $\bar K^0$ production, there is a significant contribution from s-channel  $\Lambda$ term, 
which flattens for a wide range of $p_k$. 
Its interference with the contact term shifts the peak for the kaon momentum distribution.

As an example, we have applied in our calculation some cuts corresponding to the KAOS spectrometer at MAMI. Following Ref.~\cite{Achenbach:2011eb},
the kaon momentum has been restricted to the range $400 - 700 \; MeV/c$ and the kaon angle to the range  $ 21 - 43^\circ $. 
For electron energies of 1.5 GeV, these cuts would reduce the signal by a factor $\approx 6$.
Additionally, taking into account the kaons  survival probability in KAOS~\cite{Achenbach:2011zza} would further 
reduce the number of events by a similar factor.
Thus, the measurement of these cross section at currently existing facilities, with 
their luminosities and detectors would require quite long runs.   
\begin{figure}
\begin{center}
\includegraphics[width=10cm]{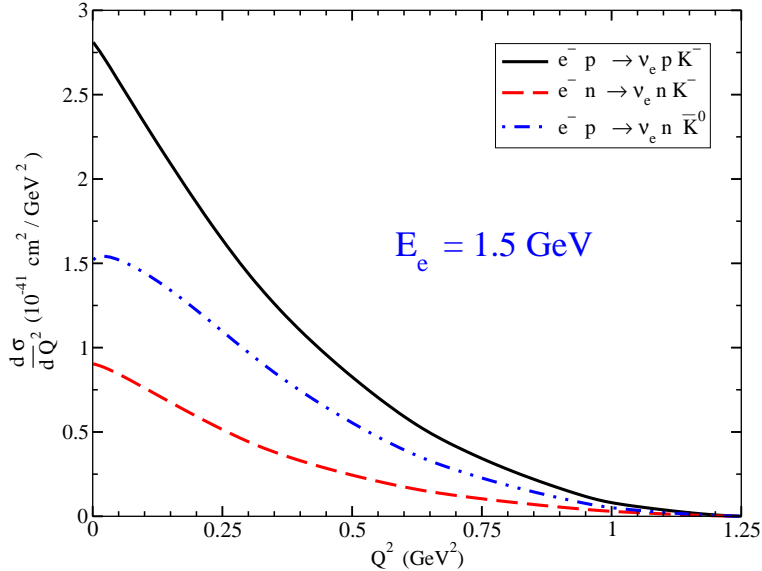}
\caption{$Q^2$ distribution at electron energy $E_e = 1.5$ GeV}
\label{fg:q2}
\end{center}
\end{figure}
 We have also investigated how the $Q^2$ dependence of the weak form factors would 
 affect our predictions. As mentioned above, very little is known for this dependence 
 given that the existing experimental information comes  from beta decay that occurs for 
 very low  $Q^2$ values. In this calculation, we are assuming a simple dipole dependence 
 and the same form factor for all background channels. Thus, we can only obtain some idea 
 on the uncertainty/sensitivity of our results to the form factors.
  As can be seen in Fig.~\ref{fg:q2}, the processes have a small contribution from large $Q^2$ values.
 Thus the size of the cross section depends moderately on the form factors.
 For instance, by changing the dipole mass($M_F = 1.05$ GeV), a 20\% up/down, 
 one gets changes of about 10\% for the neutron channel and about 28\% for the two proton channels at $E_e=1.5$ GeV.
 We have also studied the sensitivity of the electron induced cross sections
 to the D and F parameters, that govern the hyperon beta decays.
 For that, we have modified the D value by a 5\% while keeping $g_A$(=D + F)=1.26 constant.
 Our results show cross sections
 that grow around 5\% for the proton processes and decrease by a similar factor for the neutron case.
 This implies that some ratios, such as $\sigma_{\bar K^0}/\sigma_{K^-}$ on deuteron 
could be a sensitive probe for these parameters.

\begin{figure}
\begin{center}
\includegraphics[width=10cm]{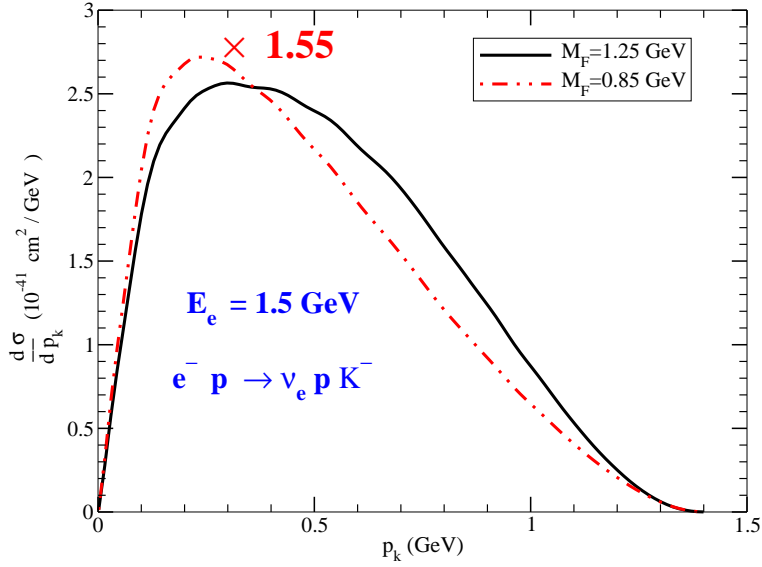}
\caption{Kaon momentum distribution at electron energy $E_e = 1.5$ GeV for the 
$ e^- + p \rightarrow \nu_e(\bar\nu_e) + K^- + p $ channel for a dipole mass 
$M_F=1.25$ and $M_F=0.85$ GeV. The second curve ($M_F=0.85$ GeV) has been scaled to get the same area.}
\label{fg:mom2}
\end{center}
\end{figure}

The measurement of the  $Q^2$ dependence would require the detection of the final nucleon in addition to the kaon. 
However, also some purely kaonic observables show some sensitivity to the form factors. For instance, Fig.~\ref{fg:mom2} 
shows how a larger dipole mass pushes the kaon  momentum towards larger values.

In summary, we have developed a microscopical model for single kaon production off nucleons induced by electrons/positrons.
This model is based on the SU(3) chiral Lagrangians. The calculations are performed up to an electron/positron energy of 1.5 GeV where
we expect this model to be quite reliable. The parameters of the model 
for the background terms are  $f_\pi$, the pion decay constant, Cabibbo's angle, 
 the proton and neutron magnetic moments and the axial vector coupling constants for the baryons octet, the $D$ and $F$ couplings.
All of them are relatively well known. To account for the $Q^2$ dependence 
we have taken a global dipole form factor as in Ref.~\cite{Alam:2012zz}.
For the electron induced process, we have also considered the contribution from the $\Sigma^*(1385)$ resonance term, 
the weak couplings for which has been obtained using SU(3) symmetry from those of the $\Delta(1232)$ resonance. 
We predict cross sections that, although small, could be measured at current experimental facilities. 
Furthermore, our results could facilitate the study of the
hyperon/nucleon weak coupling constants and their form factors.

We should remark that for some of the studied channels and for the considered energies,
there are no electromagnetic competing processes that could hinder their investigation, and therefore, the present study of the
 single kaon production cross section induced by electrons/positrons opens an interesting window for research for strangeness physics
 in the weak interaction sector.
\appendix{\bf APPENDIX}

\section{Hadronic Currents}\label{app:amplitude}

\begin{table*}[ht]
 \begin{center}
\begin{tabular}{|l|c|c|c|c|c|c|c|c|c|c|} \hline \hline
	    Process   	  	       &$B_{CT}$&$A_{CT}$& $A_{\Sigma}$ &$A_{\Lambda}$&$A_{Cr\Sigma}$&$A_{Cr\Lambda}$&$A_{KP}$& $A_{\pi }$ & $A_{\eta }$ & $ A_{\Sigma^*} $ \\\hline		   
$ e^- n \rightarrow \nu  K^- n        $&    D-F &    1   &    -1	&   0	      &    0	     &    0	    &-1       &    1	   &   1	 &  2		    \\		
$ e^- p \rightarrow \nu  K^- p        $&     -F &   2	 &$-\frac{1}{2}$&   1	      &    0	     &    0	    &-2       &   -1	   &   1	 &  1		    \\		
$ e^- p \rightarrow \nu \bar K^0 n    $&   -D-F &   1	 & $\frac{1}{2}$&   1	      &    0	     &    0	    &-1       &   -2	   &   0	 &  -1  	    \\		
$ e^+ n \rightarrow \bar\nu  K^+ n    $&    D-F &   -1   &    0 	&    0        &   -1	     &    0	    &-1       &    -1	   &   -1	 &  0		    \\  	     
$ e^+ p \rightarrow \bar\nu  K^+ p    $&     -F &   -2	 &    0 	&    0        &$-\frac{1}{2}$&    1	    &-2       &    1	   &   -1	 &  0		    \\  	     
$ e^+ n \rightarrow \bar\nu  K^0 n    $&   -D-F &   -1	 &    0 	&    0        & $\frac{1}{2}$&    1	    &-1       &    2	   &   0	 &  0		    \\ \hline \hline   
 \end{tabular}
\caption{Constant factors  appearing in the hadronic current}\label{tb:currents}
 \end{center}
\end{table*}

The contributions to the hadronic current are 

\begin{eqnarray*}
J^\mu \arrowvert_{CT} &=&i A_{CT} V_{us} \frac{ \sqrt{2}}{2 f_\pi}  \bar N(p^\prime) \; (\gamma^\mu + B_{CT} \; \gamma^\mu \gamma_5 ) \; N(p) \\
j^{\mu}\big|_{Cr\Sigma} &=& i A_{Cr\Sigma} V_{us} \frac{\sqrt{2}}{2 f_\pi} \bar N(p^\prime)
 \left( \gamma^\mu 
+i\frac{\mu_p+2\mu_n}{2M}\sigma^{\mu\nu}q_\nu
+ (D-F)(\gamma^\mu-\frac{q^\mu}{q^2-M_k^2}\q )\gamma^5 \right)\nonumber\\
& &\times  \frac{\p - \p_k + M_\Sigma}{( p -  p_k)^2 -M_\Sigma^2} \p_k \gamma^5  N(p)  ,\nonumber\\
j^{\mu}\big|_{Cr\Lambda}&=&i A_{Cr\Lambda} V_{us} \frac{\sqrt{2}}{4 f_\pi} \bar N(p^\prime)\left( \gamma^\mu
 +i\frac{\mu_p}{2M}\sigma^{\mu\nu}q_\nu
-\frac{D+3F}{3} (\gamma^\mu -\frac{q^\mu}{q^2-M_k^2}\q )\gamma^5  \right)\nonumber\\
& &\times \frac{\p - \p_k +M_\Lambda}{( p -  p_k)^2 -M_\Lambda^2}   \p_k \gamma^5 N(p) ,\nonumber \\
J^\mu \arrowvert_{\Sigma} &=&i A_{\Sigma} (D-F) V_{us} \frac{ \sqrt{2}}{2 f_\pi} \bar N(p^\prime) p_k\hspace{-.9em}/ \; \gamma_5  \frac{ p\hspace{-.5em}/ +
  q\hspace{-.5em}/ + M_\Sigma}
  {( p +  q)^2 -M_\Sigma^2} \left(\gamma^\mu +i  \frac{(\mu_p + 2\mu_n)}{2 M} \sigma^{\mu \nu} q_\nu \right. \\
&+& \left. (D-F) \left\{ \gamma^\mu 
  - \frac{q^\mu}{ q^2-{M_k}^2 } q\hspace{-.5em}/ \right\} \gamma^5 \right) N(p) \\
J^\mu \arrowvert_{\Lambda} &=&  i A_{\Lambda} V_{us} (D+3F)  \frac{1} {2 \sqrt{2} f_\pi}   \bar N(p^\prime) p_k\hspace{-.9em}/ \; \gamma^5 \frac{ p\hspace{-.5em}/ +
  q\hspace{-.5em}/ +M_\Lambda}
  {( p +  q)^2 -M_\Lambda^2} \left(\gamma^\mu +i \frac{\mu_p}{2 M}  \sigma^{\mu \nu} q_\nu \right. \\
 &-& \left. \frac{(D + 3 F)}{3} \left\lbrace \gamma^\mu   - \frac{q^\mu }{ q^2-{M_k}^2 } 
q\hspace{-.5em}/ \right\rbrace \gamma^5 \right) N(p) \\
J^\mu \arrowvert_{KP}&=& i A_{KP} V_{us} \frac{\sqrt{2}}{2 f_\pi}  \bar N(p^\prime)  q\hspace{-.5em}/ \; N(p) \frac{q^\mu}{q^2-M_k^2}    \\
J^\mu \arrowvert_{\pi} &=& iA_{\pi } \frac{M\sqrt{2}}{2 f_\pi}  V_{us}  (D + F)\frac{ 2 {p_k}^\mu -q^\mu}{(q-p_k)^2 - {m_\pi}^2} \bar N(p^\prime)  \gamma_5  N(p) \\
J^\mu \arrowvert_{\eta} &=&i A_{\eta } \frac{M\sqrt{2}}{2 f_\pi}  V_{us}  (D - 3 F)\frac{2 {p_k}^\mu - q^\mu}{(q-p_k)^2 - {m_\eta}^2} \bar N(p^\prime) 
        \gamma_5 N(p) \\
J^\mu \arrowvert_{\Sigma^*} &=&- i A_{\Sigma^*} \frac{\cal C}{ f_\pi } \frac{1}{\sqrt{6}} \; V_{us} \;  
  \frac{p_k^\lambda}{P^2 - M_{\Sigma^*}^2 + i \Gamma_{\Sigma^*} M_{\Sigma^*}}\; 
\bar N(p^\prime) P_{RS_{\lambda \rho}} ( \Gamma_V^{\rho \mu} +\Gamma_A^{\rho \mu} ) N(p) 
\end{eqnarray*}
In $\Gamma_V^{\rho \mu} +\Gamma_A^{\rho \mu}$, the form factors are taken as for the $\Delta^+$ case in Ref.~\cite{Alam:2012zz}.
The extra factors for each of the $\Sigma^*$ channels are given by $A_{\Sigma^*} $ in Tab. \ref{tb:currents}.

\begin{acknowledgments}
We acknowledge  useful discussions with P. Achenbach and  S. K. Singh. M. R. A. wishes
to acknowledge the financial support from the University of Valencia and Aligarh Muslim
University under the academic exchange program and also to the C.S.I.R., Govt. of India for S.R.F. 
This work has been partially supported by the DST, Government of India under the grant SR/S2/HEP-0001/2008, 
the Spanish Ministerio de Econom\'\i a y Competitividad and European FEDER funds under
the contracts FIS2011-28853-C02-01 and FIS2011-24149, by Generalitat Valenciana under contract PROMETEO/2009/0090,
by Junta de Andalucia  grant FQM-225 and by the EU Hadron-Physics2 project, grant agreement no. 227431. I. R. S. also wishes to acknowledge the financial support from MIUR grant
PRIN-2009TWL3MX.
\end{acknowledgments}


\end{document}